\documentclass[pra,twocolumn,aps,showpacs,preprintnumbers,amsmath,amssymb,superscriptaddress]{revtex4-1}

\usepackage{braket}
\usepackage{verbatim}
\usepackage{color}
\usepackage{ulem}
\usepackage{amsmath}
\usepackage{graphicx}
\usepackage{dcolumn}
\usepackage{bm}

\begin{document}


\title{Relativistic corrections to photonic entangled states for the space-based quantum network}

\author{Ebubechukwu O. Ilo-Okeke}
\affiliation{New York University Shanghai, 1555 Century Ave, Pudong, Shanghai 200122, China}  
\affiliation{Department of Physics, School of Physical Sciences, Federal University of Technology, P. M. B. 1526, Owerri 460001, Nigeria}
 
\author{Batyr Ilyas}
\affiliation{Department of Physics, Nazarbayev University, 53 Kabanbay Batyr Ave., Astana 0100006  Kazakhstan}  
\affiliation{Department of Physics, Massachusetts Institute of Technology, Cambridge MA 02139 USA}

\author{Louis Tessler}
\affiliation{New York University Shanghai, 1555 Century Ave, Pudong, Shanghai 200122, China} 
\affiliation{Department of Physics and Astronomy, Macquarie University, Sydney, NSW, 2109, Australia} 

\author{Masahiro Takeoka}
\affiliation{National Institute of Information and Communications Technology, Koganei, Tokyo 184-8795, Japan} 

\author{Segar Jambulingam}
\affiliation{New York University Shanghai, 1555 Century Ave, Pudong, Shanghai 200122, China}  
\affiliation{Department of Physics, Ramakrishna Mission Vivekananda College, Mylapore, Chennai 600004, India} 

\author{Jonathan P. Dowling}
\affiliation{Hearne Institute for Theoretical Physics, Department of Physics and Astronomy,Louisiana State University, Baton Rouge, Louisiana 70803, USA}
\affiliation{CAS-Alibaba Quantum Computing Laboratory, USTC, Shanghai 201315, China}
\affiliation{NYU-ECNU Institute of Physics at NYU Shanghai, 3663 Zhongshan Road North, Shanghai 200062, China}
\affiliation{National Institute of Information and Communications Technology, 4-2-1, Nukui-Kitamachi, Koganei, Tokyo 184-8795, Japan}

\author{Tim Byrnes}
\email{tim.byrnes@nyu.edu}
\affiliation{State Key Laboratory of Precision Spectroscopy, School of Physical and Material Sciences,
East China Normal University, Shanghai 200062, China}
\affiliation{New York University Shanghai, 1555 Century Ave, Pudong, Shanghai 200122, China} 
\affiliation{NYU-ECNU Institute of Physics at NYU Shanghai, 3663 Zhongshan Road North, Shanghai 200062, China}
\affiliation{National Institute of Informatics, 2-1-2 Hitotsubashi, Chiyoda-ku, Tokyo 101-8430, Japan}
\affiliation{Department of Physics, New York University, New York, NY 10003, USA}

\date{\today}

\begin{abstract}
In recent years there has been a great deal of focus on a globe-spanning quantum network, including linked satellites for applications ranging from quantum key distribution to distributed sensors and clocks. In many of these schemes, relativistic transformations may have deleterious effects on the purity of the distributed entangled pairs. In this paper, we make a comparison of several entanglement distribution schemes in the context of special relativity.  We consider three types of entangled photons states: polarization, single photon, and Laguerre-Gauss mode entangled states.  
All three types of entangled states suffer relativistic corrections, albeit in different ways.   These relativistic effects become important in the context of applications such as quantum clock synchronization where high fidelity entanglement distribution are required. 
\end{abstract}


\maketitle


\section{Introduction}

A major roadblock facing the widespread utilization of quantum communication such as quantum cryptography is the difficulty of producing long-distance entanglement.  Photons are a natural 
way of generating such entanglement due to their excellent coherence properties and the fact that they are ``flying qubits''.  However optical fiber quantum communication is limited to distances of approximately $ \sim $ 100 km due to photon loss, which make them practical for only for a limited region and not a global scale.  Broadly speaking, two approaches have been considered to overcome this challenge --  the use of quantum repeaters to cascade entanglement generation for longer distances \cite{briegel98,sangouard11}, and free-space schemes \cite{ursin07,ma12,yin12}. Recent experiments demonstrating ground-to-space entanglement distribution over 1000km \cite{yin2017satellite,ren2017ground} shows the effectiveness of space-based entanglement distribution for long distance quantum communication \cite{rarity02,kaltenbaek04,armengol08,villoresi08,xin11,rideout12,wang13,yin13,jennewein14,vallone15,tang16,carrasco16,gibney16,oi17}.  Quantum communication in space is attractive due to the negligible effects of the atmosphere, which is the origin of decoherence effects such as photon loss and dephasing.  The space-based protocol allows for the possibility of globe-scale quantum network where the photons can be transmitted at distances of the order of the diameter of the Earth without the need of additional infrastructure such as quantum repeaters.  

Some of the suggested potential applications of the space-based quantum network are quantum cryptography, clock synchronization, quantum metrology, quantum distributed computing, quantum teleportation, quantum simulation, and super-dense coding  \cite{lin,friis,ahmadi,perseguers13}. It also provides opportunities to examine fundamental physics experiments combining quantum mechanics and general relativity \cite{ralph09,ralph14,joshi17}.   In particular, clock synchronizations methods based on shared entangled states have been of particular interest since they possess distinct advantages over classical schemes \cite{jozsa00,komar,yurtsever}.  For example, in quantum clock synchronization schemes such as given in Ref. \cite{jozsa00}, once the entanglement is established, the effects of the intervening medium (i.e. the atmosphere or even the Earth itself) has no effect on the synchronization itself.  The realization of such schemes has been hindered by both theoretical and experimental difficulties.  For example, it was pointed out in Ref. \cite{preskill2000} that there is a hidden assumption of a common phase reference due to the need of performing a phase sensitive measurement at both Alice and Bob. Recently, several advances have been made both on the theoretical \cite{ilo2018remote} and experimental fronts \cite{yin2017satellite,ren2017ground} which makes the protocol more viable. In view of atomic clocks on satellites having a precision of $ 10^{-13} $, work towards improving this to  $ 10^{-15} $ is in progress \cite{schiller12,lezius2016space}, and ground-based optical atomic clocks reaching $ 10^{-18} $ and beyond \cite{ludlow2015optical}, such synchronizations require extreme accuracy and would need to incorporate relativistic effects.  This is an analogous situation to the Global Positioning System (GPS), where relativistic effects due to both special and general relativity must be accounted for.

\begin{figure}
\includegraphics[width=\columnwidth]{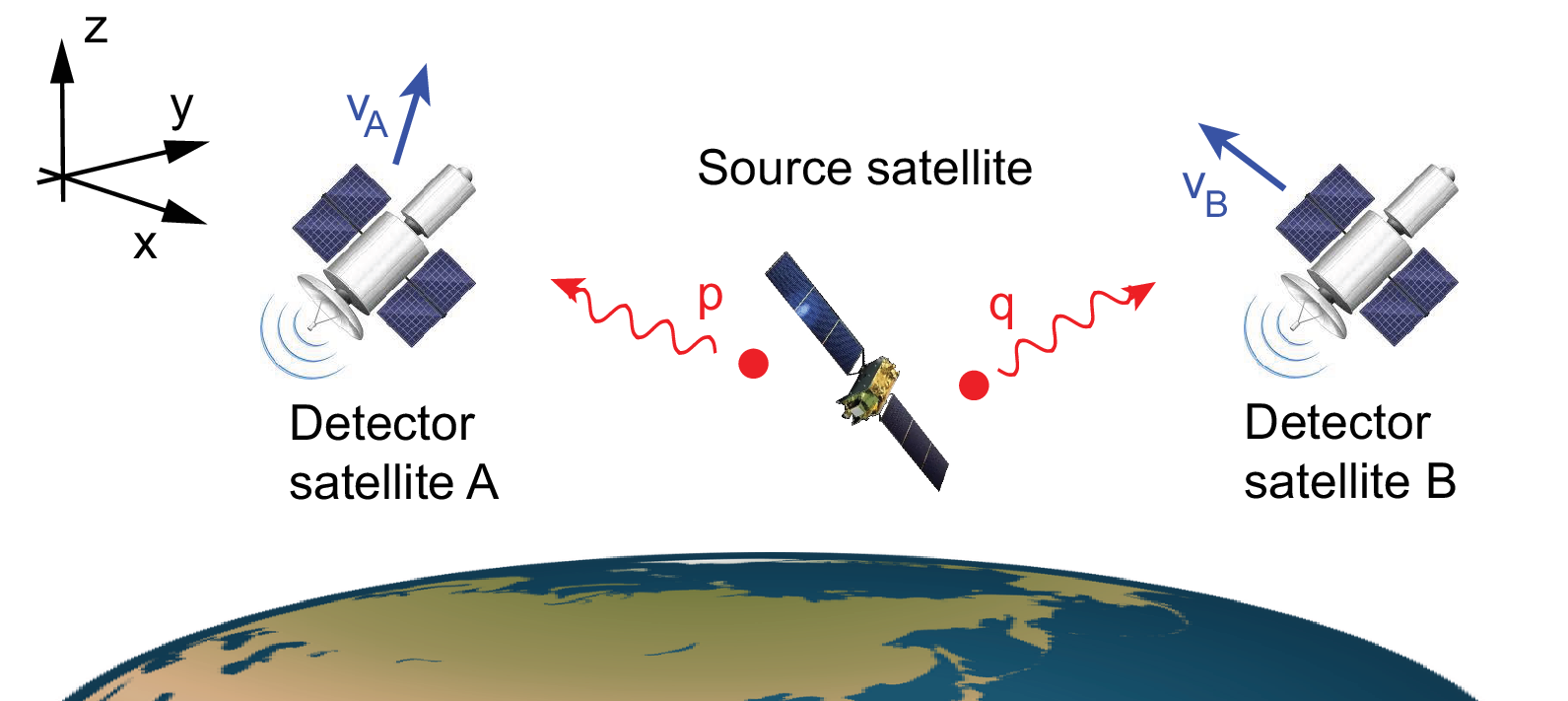}
\caption{Entanglement distribution between three satellites in LEO.  The source satellite produces entangled photons as shown in the text.  The detector satellites are moving with respect to the source satellite and each other. The photons heading to the two satellites may have different momenta $\bm{p},\bm{q} $, due to their different directions.  We choose Alice's satellite to be moving in the $ z $-direction without loss of generality. }
\label{fig1}
\end{figure}

In this paper, we investigate various strategies for space-based entanglement distribution using photons in relation to special relativistic effects.   Specifically,  we consider the satellite-to-satellite entanglement distribution configuration as shown in Fig. \ref{fig1} where the effects of the atmosphere can be largely neglected.  It has been known for some time that relativistic effects have an influence upon entanglement \cite{adami,lihui,ren}.   For instance in Ref. \cite{adami}, it was shown that amount of entanglement may change when viewed from different frames for polarization-encoded photon pairs, due to the polarization not being a Lorentz invariant (LI) quantity. 
However, this is not the only choice for entanglement distribution.  Other popular alternatives for entanglement generation include single photon entangled states and dual-rail entangled photons.  Different states respond differently under Lorentz transformations, and may be more advantageous in the context of space-based entanglement distribution.   We specifically compare the three states  (I) polarization entangled photons; (II) a single photon entangled state; and (III) Laguerre-Gauss entangled photons.  The effects of Lorentz transformations in the context of low Earth orbit (LEO) satellites producing and detecting the photons will be investigated. We will be interested specifically in how much the states change as measured by the trace distance, and effects on entanglement.

\section{Entangled states}

Let us first define the three types of entangled photon states that will be analyzed in this paper for creating long-distance entanglement using photons. The first is simply a polarization entangled photon pair, produced for example by parametric down conversion.  The state is written
\begin{align}
| \Psi_{\text{I}}^{(S)} \rangle = \frac{1}{\sqrt{2}} \left( | \bm{p}, h \rangle | \bm{q}, h \rangle - | \bm{p}, v \rangle | \bm{q} , v \rangle \right), \label{polarizationepr}
\end{align}
where $ | \bm{p}, \sigma \rangle $ is a single photon eigenstate of four momentum operator with polarization $ \sigma = h,v $, and the $ S $ refers to the fact that Alice and Bob's photons are in the reference frame of the source satellite. Throughout this paper we assume Alice's photon has momentum $ \bm{p} $ and Bob's has momentum $ \bm{q} $.  

The second type of entangled state is the single photon entangled state, which can be produced by a single photon source mounted on the source satellite entering a 50:50 beamsplitter.  The state is 
\begin{equation}
|\Psi_{\text{II}}^{(S)} \rangle = \frac{1}{\sqrt{2}}(| \bm{p}, \lambda \rangle |0 \rangle -|0\rangle | \bm{q}, \lambda \rangle).
\label{singlephotonepr}
\end{equation}
where  $ \lambda =\pm 1 $ labels the helicity, and $ \ket{0} $ is the electromagnetic vacuum. 
The entanglement here is in terms of the photon number space and the vacuum is used to encode one of the logical states.   The meaning of the entanglement in this state is that if the photon travels to Alice, then the state at Bob will be the vacuum, and vice versa.  

Finally, the third type of entangled state is using modes defined by Laguerre-Gauss modes
\begin{align}
\lvert \Psi_\mathrm{III}^{(S)} \rangle = &  \frac{1}{\sqrt{2}} \Big[ |\bm{p}, m = 1, \lambda \rangle |\bm{q}, m = 1, \lambda \rangle \nonumber \\
& - |\bm{p}, m = -1, \lambda \rangle | \bm{q}, m =-1, \lambda \rangle \Big] 
\label{typeiiistate}
\end{align}
where
\begin{align}
|\bm{p}, m ,\lambda \rangle = \int \tilde{d \bm{p}'}  f_{m}^{\bm{p}} ( \bm{p}')  | \bm{p}', \lambda \rangle .
\end{align}
where $\tilde{d\bm{p}}\equiv\frac{d^3\bm{p}}{2|\bm{p}|} $ is a Lorentz-invariant momentum integration measure. We shall consider the Laguerre-Gauss function of radial index $0$, azimuthal index $m = \pm 1$, and beam waist $w_0$ at the focal point is written in real space as \cite{andrews2013}
\begin{align}
\label{eq:li01}
f_{m}^{\bm{p}_0}  (r ,\phi ,z) 
= \frac{2 r}{\sqrt{\pi} w^2}
e^{-\frac{r^2}{w^2}} e^{-i \frac{p_0}{\hbar} \frac{r^2 z}{2\left(z^2_R + z^2\right)} } e^{i\,m\phi} e^{-i2\chi(z)},
\end{align}
where $ \bm{p}_0 = p_0 \hat{\bm{z}} $ for a given photon momentum $p_0$,
\begin{align}
w^2 & = w^2_0\left[1 + \left(\frac{z}{z_R}\right)^2 \right], \label{wfunction} \\
z_R & = \frac{\pi w_0^2}{\lambda},\\
\chi(z) &= \arctan\left(\frac{z}{z_R} \right),
\end{align}
and $ \lambda $ is the wavelength.    
The above distribution is for the case that the overall propagation direction is in the $z $-direction with the photon momentum $ p_0 $.  For other directions, a transformation of coordinates is required to obtain the final distribution. 

This can be Fourier transformed to momentum space to give the expression
\begin{equation}
\label{eq:li15}
f_{m}^{\bm{p}_0}   (r_{\bm{p}} ,\phi_{\bm{p}} ,z_{\bm{p}}) = \frac{\mathcal{N}}{\sqrt{\pi}} \frac{r_{\bm{p}}^2 w_0^3}{4} e^{i m \phi_{\bm{p}}}  e^{-\frac{r_{\bm{p}}^2 w^2_0}{4}} \delta(z_{\bm{p}}  - z_{\bm{p}}'),
\end{equation}
where  $r_{\bm{p}},\phi_{\bm{p}},z_{\bm{p}} $ is the radial, azimuthal, and longitudinal components of the momentum.  Here the radial momentum is defined within the paraxial approximation such that $z_{\bm{p}}' = p_0 - \tfrac{r_{\bm{p}}^2}{2 p_0}$. $\mathcal{N}$ is normalization constant introduced such that $\int \tilde{d \mathbf{p}} \lvert f_m^{\bm{p}_0} (\mathbf{p})\rvert^2 = 1. $  Eq. (\ref{typeiiistate}) can be thought as being a realization of a dual rail entanglement, where the modes are defined in 
terms of Laguerre-Gauss modes with quantum number $ m $.


The type I, II, III states have a different behavior under Lorentz transformations, and our task will be analyze their properties and see if there is a preferable way of performing entanglement distribution.  We note that the above three types of entangled states are not the only ones that can be realized.  For example, dual rail entanglement could be realized also using two spatially separated modes.  However, such a state is problematic in terms of long-distance entanglement distribution because the spatial modes will start to overlap due to diffraction.  In this sense the Laguerre-Gauss modes are preferable since the distinction between the modes are 
preserved even after the modes are diffracted~\cite{zhang2016}.  We have chosen the above states as three particularly interesting states which may be compatible with space-based entanglement distribution.

\section{Lorentz boost of a single photon}

\subsection{Transformation of states}

First, let us examine how single photon states transform. Momentum-helicity eigenstates in the Source frame are defined as \cite{adami,alsing}
\begin{align}
 | \bm{p}, \lambda \rangle & = R(\hat{\bm{p}}) (0, 1, i \lambda, 0 )^{T}/ \sqrt{2} ,
\end{align}
where the rotation matrix is 
\begin{align}
R(\hat{\bm{p}} ) = R_z ( \phi_{\bm{p}} ) R_y (\theta_{\bm{p}} ).  
\end{align}
Here $ R_{y,z} $ are the standard SO(3) rotation matrices, and 
$ \hat{\bm{p}} = (\sin \theta_{\bm{p}} \cos \phi_{\bm{p}}, \sin \theta_{\bm{p}} \sin \phi_{\bm{p}}, \cos \theta_{\bm{p}}) $ is the normalized 3-momentum which specifies the photon's direction. Horizontally and vertically polarized photons are defined as 
 \begin{align}
 | \bm{p},h \rangle & = R(\hat{\bm{p}}) (0, \cos \phi_{\bm{p}}, - \sin \phi_{\bm{p}}, 0 )^{T} \nonumber \\
& = \frac{1}{\sqrt{2}} ( e^{i\phi_{\bm{p}}}   
| \bm{p}, \lambda=+1 \rangle + e^{-i\phi_{\bm{p}}} | \bm{p}, \lambda=-1 \rangle ) ,  \label{hpoldef} \\
 | \bm{p},v \rangle & = R(\hat{\bm{p}}) (0, \sin \phi_{\bm{p}} , \cos \phi_{\bm{p}} , 0 )^{T} \nonumber \\
&  = \frac{-i}{\sqrt{2}} ( e^{i\phi_{\bm{p}}}  
| \bm{p}, \lambda=+1 \rangle - e^{-i\phi_{\bm{p}}} | \bm{p}, \lambda=-1 \rangle )  .  \label{vpoldef}
\end{align}
For a photon of helicity $ \lambda $ and momentum $ \bm{p} $ in the Source frame, the state in another frame is transformed as \cite{adami}
\begin{align}
U(\Lambda) | \bm{p}, \lambda \rangle = e^{-i \lambda \Theta (\Lambda, \bm{p})} | \Lambda \bm{p}, \lambda \rangle
\label{generalphotontrans}
\end{align}
where $ \Theta $ is the Wigner phase, and $ \Lambda $ is the Lorentz transformation which may include boosts and rotations. 
The Wigner phase associated for a pure rotation $ \Lambda = R(\hat{\bm{v}}) $  (no boosts)  is \cite{adami}
\begin{align}
 \Theta & ( R(\hat{\bm{v}} ) , \bm{p})  = \phi_{\bm{v}} \nonumber \\
&  + \arg \Big(\sin \theta_{\bm{v}} \cos \theta_{\bm{p}} \cos \phi_{\bm{p}} + \cos \theta_{\bm{v}} \sin \theta_{\bm{p}}  + i \sin \theta_{\bm{v}} \sin \phi_{\bm{p}} \Big) .
\end{align}
Meanwhile for a pure boost in the $ z $-direction   $ \Lambda = L_z (\beta) $ there is zero Wigner phase:
\begin{align}
 \Theta & ( L_z (\beta) , \bm{p})  = 0  .
\label{lzwignerphase}
\end{align}
The origin of the Wigner phase can be understood to be due to a rotation of the coordinate system, which induces a phase in the overall state.  Similar effects are seen in qubit systems with a redefinition of the coordinate system \cite{ilo2018remote}.  

Since Alice's satellite can be moving in any direction and is moving with respect to the Source satellite, it will in general include both boosts and rotations. The transformation  from the Source to Alice can be written as
\begin{align}
 \Lambda_{A} = R(\hat{\bm{v}}_A ) L_z (\beta) R^{-1} (\hat{\bm{v}}_A )
\label{lambdaas}
\end{align}
where $ \beta = v_A /c $  (where $c $ is the speed of light) is the velocity of Alice's satellite with respect to the Source, and $ \hat{\bm{v}}_A $ is the direction of the relative velocity of Alice with respect to the Source, and $L_z$  is a Lorentz boost along the $ z $-direction.  
Applying (\ref{lambdaas}) to (\ref{generalphotontrans}) we find that 
\begin{align}
U ( \Lambda_{A} ) | \bm{p}, \lambda \rangle =  e^{-i \lambda  \Omega (\Lambda_{A}, \bm{p})  }  | \Lambda_{A} \bm{p}, \lambda \rangle ,
\label{lambdaastransformed}
\end{align}
where
\begin{align}
\Omega (\Lambda_{A}, \bm{p})  =   \Theta (R(\hat{\bm{v}}_A ), L_z (\beta) R^{-1} (\hat{\bm{v}}_A )  \bm{p}) + \Theta (R^{-1} (\hat{\bm{v}}_A ), \bm{p})  .
\label{omegadefinition}
\end{align}
This shows that firstly the momentum of the photon is changed from $  \bm{p} $ to $ \Lambda_{A} \bm{p} $.  Due to the boost, this will involve a change of the magnitude of the momentum, corresponding to either a red- or blue-shift of the photon.  In general, the direction will also change due to Lorentz contraction of space-time.  We note that due to (\ref{lambdaas}) involving two rotations, Alice will see the Source's coordinates Lorentz contracted due to their relative motion. 
We evaluate that for a boost in an arbitrary direction (\ref{omegadefinition}) the total Wigner phase is
\begin{equation}
\label{generalwignerphase}
\Omega (\Lambda_{A}, \bm{p}) = \arg{(C + i D)}
\end{equation}
where
\begin{align}
D & =\sin(\phi_{\bm{v}} - \phi_{\bm{p}} )\sin\theta_{\bm{v}} \nonumber \\
&  \times  \left( \cos\theta_{\bm{v}}  \cosh\xi-\cos\theta_{\bm{v}}  +   \cos\theta_{\bm{p}}  \sinh\xi \right) , \\
C & = \cos(\phi_{\bm{v}}  -\phi_{\bm{p}} )\sinh\xi \sin\theta_{\bm{v}}  +\cosh\xi \sin^2\theta_{\bm{v}} \sin\theta_{\bm{p}}  \nonumber \\
&  + \cos(\phi_{\bm{v}} - \phi_{\bm{p}}) \sinh^2(\xi/2) \cos\theta_{\bm{p}} \sin 2 \theta_{\bm{v}} + \cos^2 \theta_{\bm{v}} \sin\theta_{\bm{p}}  ,
\end{align}
where $\sinh\xi = \gamma\beta $, $\cosh\xi = \gamma$, $\sinh^2\xi/2 = \gamma^2\beta^2/[2(1 + \gamma)]$, and $ \gamma = 1/\sqrt{1-\beta^2} $.  

For $ \beta \ll 1 $ we can expand the above to obtain a linear approximation to the total Wigner phase
\begin{align}
\Omega (\Lambda_{A}, \bm{p}) \approx & \arg \Big[  \sin \theta_{\bm{p}} 
+ \beta \sin\theta_{\bm{v}}  \cos ( \theta_{\bm{p}}  -\theta_{\bm{v}} )  \nonumber \\ & 
 - i \beta \cos \theta_{\bm{p}}
\sin \theta_{\bm{v}} \sin ( \phi_{\bm{p}} - \phi_{\bm{v}} ) \Big] .
\label{smallbetaomegaapprox}
\end{align}
In the limiting case of Alice's frame moving in the $ x,y,z $-directions or with zero boost $ \beta = 0  $, the total Wigner phase simplifies without approximation to
\begin{align}
 \Omega & (\Lambda_{A}, \bm{p})  = \nonumber \\
&  \left\{
\begin{array}{ll}
\arg [  \beta \cos \phi_{\bm{p}} + \sin \theta_{\bm{p}} -i \beta  \cos \theta_{\bm{p}} \sin \phi_{\bm{p}} ]   & \hspace{5mm} \hat{\bm{v}}_A = \hat{\bm{x}} \\
\arg [  \beta \sin \phi_{\bm{p}} + \sin \theta_{\bm{p}} +i \beta  \cos \theta_{\bm{p}} \cos \phi_{\bm{p}} ]  & \hspace{5mm} \hat{\bm{v}}_A = \hat{\bm{y}} \\
0  & \hspace{5mm} \hat{\bm{v}}_A = \hat{\bm{z}} \\
0  & \hspace{5mm} \beta = 0 
\end{array}
\right. .
\label{wignertotalspecial}
\end{align}
Meanwhile, for horizontally and vertically polarized light, the states transform as
\begin{align}
U ( \Lambda_{A} ) | \bm{p}, h \rangle  =  & \frac{1}{\sqrt{2}} ( e^{i \Phi_{\bm{p}}  }  
| \Lambda_{A} \bm{p}, \lambda=+1 \rangle  \nonumber \\
& + e^{-i \Phi_{\bm{p}}  }  | \Lambda_{A} \bm{p}, \lambda=-1 \rangle ) \nonumber \\
U ( \Lambda_{A} ) | \bm{p}, v \rangle  = &  \frac{-i}{\sqrt{2}} ( e^{i \Phi_{\bm{p}}  }  
| \Lambda_{A} \bm{p}, \lambda=+1 \rangle  \nonumber \\
& - e^{-i \Phi_{\bm{p}}  }  | \Lambda_{A} \bm{p}, \lambda=-1 \rangle ) 
\label{polrotu}
\end{align}
where 
\begin{align}
\Phi_{\bm{p}} = \phi_{\bm{p}} -\Omega (\Lambda_{A}, \bm{p}) .
\end{align}


If the coordinates that are used by the Source can be freely chosen, taking the $ z $-axis to be in the same direction as Alice's relative motion $ \bm{v}_A $ simplifies the algebra considerably.  In this case $ \Lambda_{A} = L_z( \beta)  $ and Alice observes the photon's direction to change according to
\begin{align}
\sin \theta_{\bm{p}} & \rightarrow \sin \theta_{\bm{p}}^{(A)}   = \frac{\sin \theta_{\bm{p}}}{\sqrt{ \sin^2 \theta_{\bm{p}} + \gamma^2 (\cos \theta_{\bm{p}} - \beta)^2}} \nonumber \\
\phi_{\bm{p}}  & \rightarrow \phi_{\bm{p}}^{(A)}   = \phi_{\bm{p}}  .
\label{relativitytrans}
\end{align}
We see that under Lorentz transformation including boosts, the momentum of the photon changes and thus will appear differently in Alice's frame. To a good approximation, for $ \beta  \ll 1 $, the variation in angle has the effect of
\begin{align}
\theta_{\bm{p}}^{(A)} \approx \pi \left( \frac{\theta_{\bm{p}}}{\pi} \right)^{1-\frac{2 }{\pi \ln 2} \beta} .
\label{thetavariation}
\end{align}
This effectively broadens or contracts the angular variation around the $ z $-axis.  The angular variation is the origin of the variation in entanglement that was observed in works such as Ref. \cite{adami}.  Meanwhile the total Wigner phase $ \Omega (\Lambda_{A}, \bm{p}) =0 $ from (\ref{wignertotalspecial}), since $ R(\hat{\bm{v}}_A) = I $.

\begin{figure}
\includegraphics[width=\columnwidth]{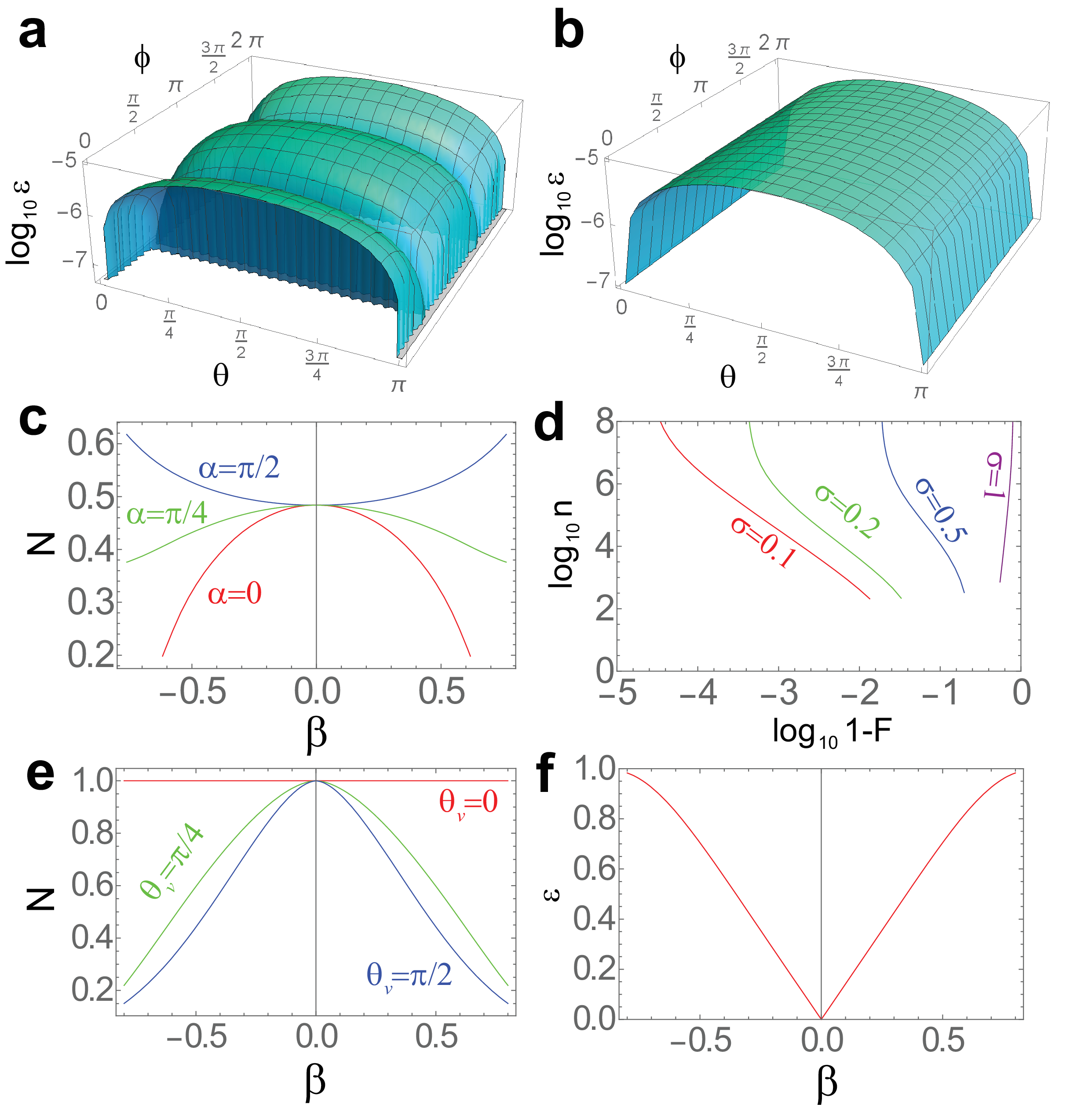}
\caption{Performance of the entanglement distribution for various protocols.  Trace distance $\varepsilon $ between the original state and that observed in a moving frame for  (a) a single horizontally (or vertically) polarized photon (b) a polarization entangled photon pair moving in opposite directions $ \theta = \theta_A = \theta_B + \pi$. The boost is in the $ z$-direction $ \Lambda_{A} = L_z (\beta) $.  Parameters are $ \beta = 10^{-5} $.  (c) Negativity of (\ref{diffractedpsi}) for type I states under a Lorentz boost in the $ z $-direction with photon directions directed along polar angle $ \alpha $ and azimuth angle $ \zeta = 0 $.  Photons are taken to move in opposite directions $ \theta_A = \theta_B + \pi$, $ \phi_A = -\phi_B $ and the spread due to the diffraction is $ \sigma = 1 $. (d) Number of entangled photon states (\ref{typeiepr}) with $ \sigma = 1 $ required to reach purities as marked.  We assume a photon attenuation factor of $ {\cal A}=100 $, and the number of photons required for $ k $ purification steps to be $ 2^k $.  (e) Negativity of (\ref{diffractedpsi}) for type II states under a Lorentz boost in the marked direction and $ \phi_{\bm{v}} = 0 $.  Photons are taken to move dominantly in the $ \hat{\bm{z}} $ and $ -\hat{\bm{z}} $-directions.  Spread due to the diffraction is $ \sigma = 1 $. (f) Trace distance of the type III state between the Source and Alice's frames under a boost in the $ x $-direction.  Only the effect of the distortion of the Laguerre-Gauss modes are considered and the Wigner phase is neglected.  }
\label{fig3}
\end{figure}

\subsection{Error introduced by Lorentz transformation}

To quantify the change, we measure the trace distance of the polarization vector
\begin{align}
\varepsilon = \text{Tr} ( \sqrt{ (\rho^{(S)} - \rho^{(A)})^2 } ) /2
\end{align}
where
\begin{align}
\rho^{(S)} & =  \text{Tr}_{\bm{p}} (| \bm{p},\sigma \rangle \langle \bm{p}, \sigma | ) \nonumber \\
\rho^{(A)} & = \text{Tr}_{\bm{p}} ( |  \Lambda  \bm{p},\sigma \rangle \langle  \Lambda  \bm{p}, \sigma | ) 
\end{align}
for this case. The momentum degrees of freedom are traced to remove any effect of the frequency and directional shift of the photons. The reason for this is that if one were to compare directly the original and boosted states we would trivially obtain $\langle  \bm{p}',\sigma  |  \bm{p},\sigma \rangle  = 0 $ for any $  \bm{p}' = \Lambda \bm{p} \ne \bm{p} $. Physically the photon in Alice's frame is either red- or blue-shifted due to the relative velocity, and hence is a different physical state.   However, if we are encoding information in degrees of freedom other than the frequency (i.e. polarization, photon number, or spatial distribution),  this fact is irrelevant since the red- or blue-shifted version of the state still retains exactly the same structure. Another way to view this is that the polarization of the red- or blue-shifted photons will be measured in the same way by the photon detectors regardless of frequency and is not relevant in terms of the encoded information.  

Fig. \ref{fig3}(a) shows the trace distance between a horizontally polarized photon with momentum $ \bm{p} $ as observed by the source and Alice's satellite moving in the $ z $-direction.  For small velocities $ \beta \ll 1 $ as will be true for all satellites orbiting the Earth, expansion of the density matrices reveals that
\begin{align}
\varepsilon_h \approx |\beta \sin \theta_{\bm{p}} \cos \phi_{\bm{p}} |,
\label{onephotonerror}
\end{align}
which very accurately summarizes the numerical results in Fig. \ref{fig3}(a).  For photons traveling along the $y $ or $ z $ axis there is no effect as horizontally polarized photons are aligned along the $ x $-axis.  We see that the basic effect of the relativistic correction on the polarization is at the level of $ \varepsilon_h  \sim O( \beta) $.  We note that the trace distance is the most appropriate quantity (than the fidelity for instance which scales as $F  \sim 1- O( \beta^2) $), as it is most closely related to distances on the Bloch sphere.  For example, in interferometric measurements, the error in the phase is proportional to the trace distance between the ideal and the state with error \cite{jozsa00}. 

\section{Lorentz boost of entangled states}

\subsection{Type I state}

Let us now examine the effect of Lorentz transformations on the entangled states.  For the type I entangled state, using (\ref{hpoldef}) and (\ref{vpoldef}) we can rewrite the original state in the Source frame in terms of helicity eigenstates
\begin{align}
| \Psi_{\text{I}}^{(S)} \rangle = & \frac{1}{\sqrt{2}} \Big( e^{i ( \phi_{\bm{p}} + \phi_{\bm{q}} )}  | \bm{p}, \lambda = +1  \rangle | \bm{q}, \lambda = +1 \rangle \nonumber \\
& - e^{-i ( \phi_{\bm{p}} + \phi_{\bm{q}} )}  
| \bm{p}, \lambda = -1  \rangle | \bm{q}, \lambda = -1 \rangle  \Big).
\end{align}
Here $ \hat{\bm{q}} = (\sin \theta_{\bm{q}} \cos \phi_{\bm{q}}, \sin \theta_{\bm{q}} \sin \phi_{\bm{q}}, \cos \theta_{\bm{q}}) $.  Transforming to Alice's frame using (\ref{lambdaastransformed}) we obtain
\begin{align}
| \Psi_{\text{I}}^{(A)} \rangle & = U(\Lambda_{A} ) | \Psi_{\text{I}}^{(S)} \rangle \nonumber \\
& = \frac{1}{\sqrt{2}} \Big( e^{i ( \Phi_{\bm{p}} + \Phi_{\bm{q}} )}  | \Lambda_{A} \bm{p}, \lambda = +1  \rangle | \Lambda_{A} \bm{q}, \lambda = +1 \rangle \nonumber \\
& - e^{-i ( \Phi_{\bm{p}} + \Phi_{\bm{q}} )}  | \Lambda_{A} \bm{p}, \lambda = -1  \rangle | \Lambda_{A} \bm{q}, \lambda = -1 \rangle  \Big), 
\label{typeiepr}
\end{align}
where $ \Phi_{\bm{p}}  = \phi_{\bm{p}} -\Omega (\Lambda_{A}, \bm{p}) $  and $ \Phi_{\bm{q}}  = \phi_{\bm{q}} -\Omega (\Lambda_{A}, \bm{q}) $.  We see explicitly that the entangled state changes under the transformation from the Source's frame to Alice's.  Firstly, the momenta of the photons are changed due to the Lorentz transformation $  \bm{p} \rightarrow \Lambda_{A}  \bm{p}$ and  $   \bm{q} \rightarrow \Lambda_{A}  \bm{q} $. This will in general introduce both a change in the direction and magnitude of the momenta. This means that the direct overlap of the states in the two frames is zero for any Lorentz transformation that is not the identity.  We note that we only consider transformations to Alice's frame since the results of transformations to Bob's frame give similar results.

For the purposes of carrying quantum information, the fact that the momenta of the photons change is 
not particularly relevant since the degrees of freedom that the entanglement is encoded is in terms of polarization or helicity.  In this case we may integrate out the momenta to write a $ 4 \times 4 $ density matrix for the states as viewed in the Source and Alice frames:
\begin{align}
\rho^{(S,A)} =  \text{Tr}_{\bm{p},\bm{q}} (| \Psi_{\text{I}}^{(S,A)} \rangle \langle \Psi_{\text{I}}^{(S,A)} | )  .
\label{rhodef}
\end{align}
For the case that Alice's motion is in the $ z $-direction, no Wigner phase is added and the sole effect to the state is the rotation of the polarization vectors, as given in (\ref{relativitytrans}).  The trace distance between the states in the Source and Alice's frames is shown in Fig. \ref{fig3}(b).  For the case of photons moving in opposite directions, the trace distance can be summarized to a very good approximation by 
\begin{align}
\varepsilon_{\text{I}} \approx |\beta \sin \theta_{\bm{p}} |.
\label{errorepr}
\end{align}
We again see that the relativistic correction again occurs at the level of $ \sim O(\beta) $. 

For satellites in LEO typically $ \beta \approx 10^{-5} $, which can be a significant effect in comparison with the precision of atomic clocks.  For example, in the clock synchronization scheme of Ref. \cite{jozsa00}, if Alice and Bob measure in different bases, this appears as an offset in the time between their clocks \footnote{For the particular scheme in Ref. \cite{jozsa00}, a suitable choice of photon bases allows the errors to reduced to $ \sim O(\beta^2) $ which scales better, but is still significant source of error.}.  One may argue that such systematic errors such as (\ref{onephotonerror}) or (\ref{errorepr}) can always be accounted for, and hence removed.  This is indeed true for GPS satellites where relativistic effects such as time dilation are compensated out.  In this way the errors could potentially be reduced to a level below  (\ref{onephotonerror}) or (\ref{errorepr}).  Then the real error estimate is then determined by how well the relativistic corrections can be corrected out, which for the case (\ref{onephotonerror}) is related to the error on the velocity estimate $ \delta \beta $.  This gives an error of $ \varepsilon \sim O( \delta \beta) $ for (\ref{errorepr}).  Since the precise velocities of the satellites are typically not known to extremely high precision, the relativistic errors can be significant, even if they are accounted for.   For example, if the velocity of the satellite is known with relative error of  $ \sim 10^{-6} $ \cite{hobbs}, thus amounts to an error $ \varepsilon \sim 10^{-11} $, which is still large in comparison to the precision of atomic clocks.   

We note that although the relativistic transformation changes the nature of the quantum state, the amount of entanglement is preserved since (\ref{typeiepr}) is also a maximally entangled state.  Thus if the task is to distribute an entangled state without any specification to the particular state, then the type I state serves this purpose.

\subsection{Type II state}

For type II single photon entangled states, transforming to Alice's reference frame, we find
\begin{align}
& |\Psi_{\text{II}}^{(A)} \rangle =    U(\Lambda_{A} ) |\Psi_{\text{II}}^{(S)} \rangle  \nonumber \\
& =   \frac{1}{\sqrt{2}}( e^{-i \lambda  \Omega (\Lambda_{A}, \bm{p})  }  |\Lambda_{A} \bm{p}, \lambda \rangle |0 \rangle -  e^{-i \lambda  \Omega (\Lambda_{A}, \bm{q})  }  |0 \rangle | \Lambda_{A} \bm{q}, \lambda \rangle).
\label{singlephotonepr2} 
\end{align}
As previously, the momenta of the photons experience a shift in magnitude and direction depending upon the direction of the boost. Although helicity is a Lorentz invariant quantity, due to the presence of the Wigner phase the entangled state can still become modified due to the transformation.   If again we trace out the momentum degrees of freedom and compare the $ 4 \times 4 $ density matrix (\ref{rhodef}) defined in the Fock spaces for the Source and Alice frames, the trace distance can be simply evaluated to be 
\begin{align}
\epsilon_{II} = \sin \frac{| \Omega (\Lambda_{A}, \bm{p}) -\Omega (\Lambda_{A}, \bm{q}) |}{2} .
\label{type2transformed}
\end{align}
Using the approximation (\ref{smallbetaomegaapprox}) we can evaluate the expression as 
\begin{align}
\epsilon_{II} \approx &  \frac{1}{2} |\beta \sin \theta_{\bm{v}}  | \nonumber \\
& \times \left| \cot \theta_{\bm{p}} \sin ( \phi_{\bm{p}} - \phi_{\bm{v}}) 
- \cot \theta_{\bm{q}} \sin ( \phi_{\bm{q}} - \phi_{\bm{v}}) \right| ,
\end{align}
which is valid for $ | \beta \cot \theta_{\bm{p}} |, | \beta \cot \theta_{\bm{q}} | \ll 1 $. We again see that the relativistic correction occurs at the level of $ \sim O(\beta) $.  The error of these states hence arises purely due to the Wigner phase between the states.  As with type I states, the entanglement is invariant under the transformation since (\ref{type2transformed}) is a maximally entangled state.  
This can be understood to be a result of the fact that type II states encode the entanglement in terms of photon number states.  From Ref. \cite{avron}, it is known that  photon number states, including the vacuum are invariant states under Lorentz transforms, and remain orthogonal in all reference frames.

\subsection{Type III state}

\label{sec:typeiiistatenodiff}

To see the effect of a boost on type III states, first let us see how one of the Laguerre-Gauss modes transforms
\begin{align}
& U(\Lambda_A ) |\bm{p}, m ,\lambda \rangle  \equiv |\bm{p}, m ,\lambda, \Lambda_A \rangle   \nonumber \\
& = \int \tilde{d \bm{p}'}  e^{-i \lambda \Omega (\Lambda_{A}, \bm{p}') } f_{m}^{\bm{p}} ( \bm{p}')  | \Lambda_{A} \bm{p}', \lambda \rangle  \nonumber \\
& = \int \tilde{d \bm{p}'}  e^{-i \lambda \Omega (\Lambda_{A},  \Lambda_{A}^{-1} \bm{p}') } f_{m}^{\bm{p}} (  \Lambda_{A}^{-1} \bm{p}')  | \bm{p}', \lambda \rangle ,
\label{transformedlgmode}
\end{align}
where we made a change of variables $ \bm{p}' \rightarrow \Lambda_{A}^{-1} \bm{p}' $ in the last line. The transformed type III state thus reads
\begin{align}
&   | \Psi_{\text{III} }^{(A)} \rangle  = U(\Lambda_A)  | \Psi_{\text{III}}^{(S)} \rangle  \nonumber \\
& = \frac{1}{\sqrt{2}} \Big(  |\bm{p}, m=1 ,\lambda, \Lambda_A \rangle  |\bm{q}, m=1 ,\lambda, \Lambda_A \rangle    \nonumber \\
& -|\bm{p}, m=-1 ,\lambda, \Lambda_A \rangle  |\bm{q}, m=-1 ,\lambda, \Lambda_A \rangle    \Big)  . 
\end{align}
This can also be explicitly be written
\begin{align}
&  | \Psi_{\text{III} }^{(A)} \rangle  = \frac{1}{\sqrt{2}} \int \tilde{d \bm{p}'} \tilde{d \bm{q}'} e^{-i \lambda ( \Omega (\Lambda_{A}, \Lambda_{A}^{-1}  \bm{p}' )  +  \Omega (\Lambda_{A}, \Lambda_{A}^{-1}  \bm{q}' ) ) }  \nonumber \\
& \times \Big[  f_{1}^{\bm{p}} ( \Lambda_{A}^{-1}   \bm{p}' )   f_{1}^{\bm{q}} ( \Lambda_{A}^{-1}  \bm{q}' )  -  f_{-1}^{\bm{p}}  (\Lambda_{A}^{-1}  \bm{p}' )   f_{-1}^{\bm{q}} (\Lambda_{A}^{-1}  \bm{q}' )  \Big] \nonumber \\
& \times  | \bm{p}' , \lambda \rangle  | \bm{q}' , \lambda \rangle  
\label{eq:li22}
\end{align}
We can see that the Laguerre-Gauss modes are not in general invariant under Lorentz transformations as they distort the momentum distribution.  As with the previous sections, we would like to see to what degree the state is preserved under a Lorentz boost which are encoded by the $ m = \pm 1 $ Laguerre-Gauss modes. 

First let us now consider the effect of the boost on a single Laguerre-Gauss mode function $ f_{m} $. To see the largest effects due to the boost, consider a Laguerre-Gauss beam that is boosted along a perpendicular direction (along the $x$-axis) to its propagation (along the $z$-axis).   
From a standard Lorentz transformation we find that 
\begin{align}
\Lambda_A^{-1} \hbar \omega_{\bm{p}} /c & = \gamma (\hbar \omega_{\bm{p}} /c - \beta x_{\bm{p}}) \nonumber \\
\Lambda_A^{-1} x_{\bm{p}} & = \gamma (x_{\bm{p}} - \beta \hbar \omega_{\bm{p}} /c) \nonumber \\
\Lambda_A^{-1} y_{\bm{p}} & = y_{\bm{p}} \nonumber \\
\Lambda_A^{-1} z_{\bm{p}} & = z_{\bm{p}} ,
\end{align}
where $ p_0 = \hbar \omega/c $. Then noting that the cylindrical polar and the Cartesian coordinate are related as $r_{\bm{p}}^2 = x_{\bm{p}}^2 + y_{\bm{p}}^2$, $\phi_{\bm{p}} = \arctan (\tfrac{y_{\bm{p}}}{x_{\bm{p}}})$, then  
\begin{align}
(\Lambda_A^{-1} r_{\bm{p}})^2 & = (\Lambda_A^{-1} x_{\bm{p}})^2 + (\Lambda_A^{-1} y_{\bm{p}})^2 \nonumber, \\
\Lambda_A^{-1} \phi_{\bm{p}} & = \arctan (\tfrac{\Lambda_A^{-1} y_{\bm{p}}}{\Lambda_A^{-1}x_{\bm{p}} }) ,
\end{align}
and 
\begin{align}
\label{eq:p01}
f_m^{ \bm{p}_0 }   (\Lambda_A^{-1} & r_{\bm{p}},\Lambda_A^{-1} \phi_{\bm{p}},z_{\bm{p}}) = \frac{\mathcal{N}}{\sqrt{\pi}} \frac{(\Lambda_A^{-1}r_{\bm{p}})^2 w_0^3}{4}   \nonumber\\
& \times  e^{i m \Lambda_A^{-1}\phi_{\bm{p}}} e^{-\frac{(\Lambda_A^{-1}r_{\bm{p}})^2 w^2_0}{4}} \delta(z_{\bm{p}} - z_{\bm{p}}'),
\end{align}
where $m= \pm 1$. Meanwhile the Wigner phase is in this case
\begin{align}
\Omega(\Lambda_A, \Lambda_A^{-1} \bm{p}) = & 
\arg [  \beta \cos \phi_{\Lambda_A^{-1} \bm{p}} + \sin \theta_{\Lambda_A^{-1} \bm{p}} \nonumber \\
& -i \beta  \cos \theta_{\Lambda_A^{-1} \bm{p}} \sin \phi_{\Lambda_A^{-1} \bm{p}} ] .
\end{align}

To calculate the error induced by the Lorentz boost, let us first consider the type of measurement that might be performed to detect the state.  From an experimental point of view, the Laguerre-Gauss modes can be detected at the single photon level by a spatial interference method \cite{leach2002}. In this technique, the Laguerre-Gauss modes are put in a mode sorter such that various angular momentum states $ m $ can be distinguished. Since the Laguerre-Gauss modes form a complete set, the transformed state (\ref{transformedlgmode}) can be expanded as a superposition of Laguerre-Gauss modes in the local frame. The trace distance between the state in the Source's and Alice's frame can be calculated from the fidelity according to
\begin{align}
\epsilon_{III} = \sqrt{ 1- | \langle \Psi_{III}^{(S)} | \Psi_{III}^{(A)} \rangle |^2 } ,
\end{align}
since both states are pure states.  The error induced by the distortion of the Laguerre-Gauss distribution due to the boost in the $ x$-direction is plotted in Fig. \ref{fig3}(f).  We see that the error in the state takes the value $ \epsilon_{III} \sim O(\beta) $ and has a linear relationship for most of the range of $ \beta $.  

While the mode sorter is the most natural method to perform a measurement of the Laguerre-Gauss modes, we point out that the distortion of the momentum (and hence spatial) distribution does not affect the topology of the overall azimuthal phase pattern, as can be seen by comparing (\ref{eq:li15}) and (\ref{eq:p01}).  Specifically, the
 winding number $ m $ of the optical phase vortex does not change under a Lorentz transformation.  
 Thus in principle the different states indexed by $ m $ should be distinguishable, as it is a topological invariant. 
This is a similar effect to how topological quantum states are unaffected by small local transformations of the states \cite{PhysRevA.92.023629}.  Such a measurement would require a measurement of the topological charge, rather than simply measuring in the local Laguerre-Gauss mode basis.  The Wigner phase that occurs due to the transformation also does not affect the relative phase of the Bell state since it acts globally on the state.  We note that these properties of Laguerre-Gauss modes have been studied with similar conclusions  in Ref. \cite{spedalieri2006quantum}.

\section{Diffraction effects}

Up to this point, we have made one idealization in that the effects of photon diffraction were not included.  In a more realistic situation, the photons will have a spread due to diffraction and will have a superposition of different momenta $ \bm{p}, \bm{q}$. All three types of states that were considered (\ref{typeiepr}), (\ref{singlephotonepr2}), (\ref{eq:li22}) in fact have the same entanglement as a maximally entangled Bell state in all frames.  As discussed in Ref. \cite{adami}, relativistic effects can affect the amount of entanglement as it changes the diffractive spread of the photons.  This type of error is of relevance to our case as it is not a systematic error that is correctable through local operations on Alice and Bob's satellites. 

We now discuss how such relativistic corrections affect the three types of photonic entangled states.  For type I and II states, to take into account of diffraction, we integrate with a momentum distribution \cite{adami}
\begin{align}
\label{diffractedpsi}
| \Psi_{\text{diff}} \rangle = \int \tilde{d \bm{p}} \tilde{d \bm{q}}  g_A ( \bm{p}) g_B (\bm{q}) | \Psi(\bm{p},\bm{q}) \rangle 
\end{align}
where the $ | \Psi (\bm{p},\bm{q}) \rangle  $ are the states (\ref{polarizationepr}) and (\ref{singlephotonepr}) in the source satellite's frame and  $g (\textbf{p})$ is a normalized diffraction function. The type III states (\ref{typeiiistate}) already have a spatial distribution and do not require integration as we explain further below. 
For a specific model of the photon spread, we follow the same form as that given in Ref. \cite{adami} where only angular spread of photons traveling in the \textit{z}-direction were considered, and the magnitude of the momentum is set to a constant. A photon traveling in arbitrary direction, obtained by rotating the photon traveling along \textit{z}-axis about \textit{y} and \textit{z} axes by the angles $\zeta$ and $\alpha$, respectively, would have a Gaussian spread about the \textit{z}-axis of the form 
\begin{equation}
g(\bm{p} )=\frac{1}{\sqrt{M}} e^{-\frac{(\theta_{\bm{p}}'')^2}{2 \sigma^2 } }  \delta(|\bm{p}|-p_0),
\end{equation} 
where $\sigma$ is a parameter controlling the angular spread of the beam and $ M $ is a suitable normalization factor, 
\begin{align}
\theta_{\bm{p}}'' & = \cos^{-1} \left(\cos \alpha \cos \theta_{\bm{p}} + \sin \alpha \sin \theta_{\bm{p}} \cos \phi_{\bm{p}} \right) \nonumber, \\
\phi_{\bm{p}}'' & = \tan^{-1} \left( \frac{A \sin\zeta  + \cos\zeta \sin \theta_{\bm{p}} \sin \phi_{\bm{p}}}{\sin\zeta \sin\phi_{\bm{p}} \sin\theta_{\bm{p}} - A \cos\zeta } \right),
\end{align}
and $A = \sin \alpha \cos \theta_{\bm{p}} -  \cos \alpha \sin \theta_{\bm{p}} \cos \phi_{\bm{p}} $. To transform to Alice's frame, one then applies a boost in the $ z $-direction to the states, which amounts to making the transformation (\ref{relativitytrans}).  

Figure \ref{fig3}(c) shows the entanglement as a function of the satellite velocity for type I photons traveling in opposite directions and various boost angles. In contrast to previous works \cite{adami}, for boosts aligned to the photon propagation ($\zeta = \alpha = \phi = 0 $), we find that the entanglement always degrades regardless of direction.  This is due to the different geometry that we consider that is relevant for our case.  For photons traveling in opposite directions, the Gaussian distribution tightens for one of the photons but broadens for the other photon according to (\ref{thetavariation}), which always results in a degradation of the entanglement. For boosts that are perpendicular to the photon propagation ($\zeta = \phi = 0,\,\alpha = \pi/2 $), the entanglement can be increased, as the Gaussian spread is redistributed towards the $ z $-axis, resulting in an effective tightening of the distribution.  

We now estimate the order to which the relativistic corrections affect the entanglement.  To gauge this we calculate the effect of the boost on the purity of the states $ P = \text{Tr} \rho^2 $. The purity is directly related to the entanglement in this case as for the case with no diffraction, the entanglement is invariant under all boosts.   The degradation in the entanglement observed in Fig.  \ref{fig3}(c) arises from an effective decoherence entering the system due to tracing out the momentum degrees of freedom.  Performing an expansion for $ \beta \ll 1 $ we find that the purity behaves as
\begin{align}
P \approx  1 - 2 \sigma^2 (1 + | \beta |)^2 .
\end{align}
As expected for no diffraction $ \sigma = 0 $, there are no relativistic corrections.  The relativistic corrections to lowest order act to accentuate the diffraction effects which are already present.  In terms of physical parameters, the diffraction angle can be estimated as $ \sigma \approx \lambda/d $, where $ \lambda $ is the photon wavelength and $ d $ is the diameter of the transmitter.  For infrared photons, this gives $ \sigma \sim 10^{-6} $.  We see that in this case the relativistic corrections are quite small as it is a secondary correction.  

Diffraction effects can be remedied using entanglement purification methods.  We demonstrate that it is possible to achieve high purities by adapting the purification procedure devised in Ref. \cite{bennett96} to our relativistic entangled photons.  In Fig. \ref{fig3}(d) we show the results of the entanglement purification on the state (\ref{rhodef}) using (\ref{diffractedpsi}).  We calculate the number of photons required as the number of photons required for a purification of a particular target fidelity, multiplied by the photon attenuation factor (the ratio of the number of photons sent to received), divided by the success probability of the purification.  The photon attenuation is $ {\cal A} = L^2 \lambda^2/d_S^2 d_A^2 $, which for parameters $ L = 13000 $ km, $ \lambda = 800 $ nm, $ d_S = d_A = 1 $ m gives $ {\cal  A} \approx 100 $ photons being sent for each one received \cite{aspelmeyer03}.  For the various diffractive spreads $ \sigma $ considered, we find that an improvement in the fidelity is achievable as long as the original diffractive spread is lower than $ \sigma \lesssim 1 $.  For very broad $ \sigma $ the purification fails and the fidelity decreases. As typically the spread is $ \sigma \ll 1 $ we anticipate that such purification methods should always be successful in practice.  

Turning to type II states, tracing out the momentum degrees of freedom we obtain an effective $ 4 \times 4 $ density matrix 
\begin{align}
\rho_{II} = \frac{1}{2} \left(
\begin{array}{cccc}
0 & 0 & 0 & 0 \\
0 & 1 & - I_{\bm{p}} I_{\bm{q}} & 0 \\
0 & - I_{\bm{p}}^* I_{\bm{q}}^* & 1 & 0 \\
0 & 0 & 0 & 0 
\end{array}
\right) .
\label{type2densitymatrix}
\end{align}
where
\begin{align}
I_{\bm{p}} = \frac{1}{M} \int \sin \theta_{\bm{p}} d \theta_{\bm{p}}   d \phi_{\bm{p}} 
e^{- \frac{(\theta_{\bm{p}}'')^2}{\sigma^2} } e^{-i \lambda \Omega( \Lambda_A, \bm{p})} .
\end{align}
The basis states of the density matrix (\ref{type2densitymatrix}) are vacuum, single photon with momentum in the vicinity of $ \bm{p} $, single photon with momentum in the vicinity of $ \bm{q} $, two photon state with momenta $ \bm{p} $ and $ \bm{q} $ respectively.  The logarithmic negativity takes a simple form for the above state, and can be written exactly
\begin{align}
N = \log_2 ( 1 +  | I_{\bm{p}} I_{\bm{q}} |^2 ) .
\end{align}

For type II states, the degradation in the entanglement comes entirely from the presence of the Wigner phase $ \Omega(\Lambda_A, \bm{p}) $.  In the limit of $ \sigma \rightarrow 0 $, the integrals $ I_{\bm{p}} \rightarrow 1 $ and the state is perfectly entangled.  In type I states,  the entanglement was degraded due to Lorentz transformations changing the spread in the values of $ \bm{p} $ and $ \bm{q} $ in (\ref{diffractedpsi}), producing a mixed state in the polarizations.  For type II states, the entanglement is encoded in the orthogonality of the Fock states. Since all the photons are of the same helicity, the mixing occurs only due to the presence of the Wigner phases.  In Fig. \ref{fig3}(e) we show the negativity for boosts in the various directions.  As expected, for a boost in the $ z $-direction, there is no change in the entanglement, since no Wigner phase is acquired.  For other directions, there is a degradation of entanglement due to incoherent mixing of various entangled state with differing Wigner phases.  We note that to see significant degradation, both large diffractive spreading $ \sigma $ and boost velocities $ \beta $ must be considered.  Since both are small parameters in practice,  as with type I states the corrections to entanglement should be quite small for this case.

One of the disadvantages of using type II state is that while Fock state measurements are naturally implemented, performing measurements in superposition bases of vacuum and single photons are more difficult than other methods. Various methods have been proposed to perform  measurements in a superposition basis of the vacuum and a single photon \cite{lee03,takeoka05,takeoka06}.  In addition to the technical overhead for performing measurements, there are additional complications for type II in overcoming photon loss.  For type I states, entanglement is encoded on the two photons that are each detected at two spatially separated locations. If one or both of the photons are lost during the transmission to Alice or Bob, then these outcomes can be safely discarded, since the two photons are not detected. In this sense the entanglement encoded in type I states are robust to photon loss. This is not the case for type II states, since a loss event does not necessarily correlate to a particular measurement outcome, due to the use of the vacuum state as one of the logical states.

In contrast to the type I and II states where we can calculate the entanglement degradation due to diffraction, in the case of type III states, it makes less sense to do such a calculation.  The reason is that for type I and II states we can still define entanglement due to photons arriving with particular momenta $ \bm{p}, \bm{q} $ in (\ref{diffractedpsi}) respectively.  However, for type III states the encoding is in the spatial mode distributions themselves.  So one cannot  integrate out the momentum degrees of freedom to obtain a mixture of different entangled states.  
Since type III states are Laguerre-Gauss modes, the spatial distribution is already given specified by the functions (\ref{eq:li01}).  The effect of long-distance communication of such photons is that the the radius $ w $ of the distribution spreads out for larger distances as given by (\ref{wfunction}).  Thus the analysis of Sec. \ref{sec:typeiiistatenodiff} still holds in this diffractive case.  The main challenge in the context of long-distance communication is that the radius $ w $ can grow very large after the photon traveling long distances.  Thus when performing the mode-sorting as given in Ref. \cite{leach2002,spedalieri2006quantum}, one must perform the rotation of the Laguerre-Gauss modes using the enlarged diffracted beam.  If the mode radius is larger than the experimental apparatus, this will contribute to photon loss.  This is also true for alternative detection methods based on measuring the topological charge \cite{PhysRevA.92.023629}.

\section{Conclusions}

In summary, we have analyzed several photon-based entanglement distribution protocols for the space-based quantum network.  We have calculated to the relativistic corrections for three types of entangled states:  polarization-based entangled states (type I),  single photon entangled states (type II), and Laguerre-Gauss entangled states (type III). The origin of the error for the type I state is that polarization is not a relativistically invariant quantity, due to the presence of the Wigner phases as given in (\ref{polrotu}).  For type II states, a Wigner phase is induced which changes the nature of the Bell state. For type III states, the Laguerre-Gauss mode is distorted due to the spatial Lorentz contractions.  We find that in terms of the trace distance of the states, all three states are affected to a level $ \sim O(\beta) $.   While in principle these are correctable if the velocities of the satellites are known to high precision, this can still introduce errors at the $ \delta \beta $, which is the error on the estimate of the satellite velocity.  In the case of zero diffraction all three states contain the maximum amount of entanglement, equal to that of a Bell state.  However, when diffraction is accounted for type I and II states degrade in entanglement since there is a mixture of various Bell states.  

 One of the most interesting applications of space-based entanglement is clock synchronization, which is currently performed using classical signals, which requires precise knowledge of the position of the satellites.  Entanglement-based methods can potentially eliminate this requirement, but as have shown in this paper, to properly take advantage of this it is important to consider the relativistic effects on the entangled states.  
In addition, the entanglement can be used for several important tasks such as 
quantum cryptography which can be used without further components such as a quantum memory.  For applications that require a quantum memory to further manipulate the entanglement, it is likely necessary to understand the relativistic effects on transfer and storage, if one requires a high fidelity protocol. Although these are beyond the scope of this paper, it is likely that these operations will generally be susceptible to relativistic effects.  To overcome this, it may be advantageous to use encoding and measurement techniques that are based on topological invariants, such as that present in type III states.  This would provide a way to overcome such relativistic corrections, since the overall topology of the quantum states would be invariant under Lorentz transformations.

\acknowledgments

T. B. is supported by the Shanghai Research Challenge Fund; New York University Global Seed Grants for Collaborative Research; National Natural Science Foundation of China (61571301,D1210036A); the NSFC Research Fund for International Young Scientists (11650110425,11850410426); NYU-ECNU Institute of Physics at NYU Shanghai; the Science and Technology Commission of Shanghai Municipality (17ZR1443600,19XD1423000); and the NSFC-RFBR Collaborative grant (81811530112).  E. O. I. O. would like to acknowledge support from the China Science and Technology Exchange Center (NGA-16-001). J. P. D. would like to acknowledge support from the US Air Force Office of Scientific Research, the Army Research Office, the National Science Foundation, and the Northrop-Grumman Corporation.


\end{document}